\documentclass{aa}
\usepackage{txfonts}
\usepackage{natbib}
\usepackage{graphicx}
\bibpunct{(}{)}{;}{a}{}{,}
\def\tbol{$T_{\rm bol}$}
\def\alf{${\alpha}$}
\def\msun{$\rm M_\odot$}
\def\lsun{$\rm L_\odot$}
\def\mdisk{$M_{\rm disk}$}
\def\menv{$M_{\rm env}$}
\def\lstar{$L_{\rm star}$}
\def\mdot{$\dot{M}$}
\def\mum{$\mu$m}
\def\percc{$\rm cm^{-3}$}
\def\persc{$\rm cm^{-2}$}
\def\h2o{$\rm H_{2}O$}
\def\co2{$\rm CO_{2}$}

\begin{document}

\title{Characterizing the nature of embedded young stellar objects through silicate, ice and millimeter observations } 


\author{A. Crapsi\inst{1,2}
    \and E. F. van Dishoeck\inst{1}  
    \and M. R. Hogerheijde\inst{1}
    \and K. M. Pontoppidan\inst{3}
    \and C.P. Dullemond\inst{4}
    } 

\authorrunning{Crapsi et al.}
\titlerunning{The true nature of class I YSO}

\offprints{A. Crapsi, \email{crapsi@strw.leidenuniv.nl}}

\institute{Sterrewacht Leiden, Leiden University, P.O. Box 9513, 2300 RA Leiden, the Netherlands
 \and Observatorio Astron\'omico Nacional (IGN), Alfonso XII, 3, E-28014 Madrid, Spain
 \and  Hubble Fellow, Division of GPS, Mail Code 150-21, California Institute of Technology, Pasadena, CA 91125, USA
 \and Max-Plank-Institut f\"{u}r Astronomie, Koenigstuhl 17, 69117 Heidelberg, Germany
 } 

\date{Received.../Accepted... }

\abstract 
{Determining the evolutionary stage of a Young Stellar Object (YSO) is of fundamental importance to test star formation theories.
Classification schemes for YSOs are based on evaluating the degree of dissipation of the  surrounding envelope, whose main effects are the extinction of the optical radiation from the central YSO and re-emission in the far--infrared to millimeter part of the electromagnetic spectrum.
Since extinction is a property of column density along the line of sight, the presence of a protoplanetary disk may lead to a misclassification of pre--main sequence stars with disks when viewed edge--on.}
 {We performed radiative transfer calculations to show the effects of different geometries on the main indicators of YSO evolutionary stage. In particular we tested not only the effects on the infrared colors, like the slope \alf \ of the flux between 2.2 and 24~\mum , but also on other popular indicators of YSO evolutionary stage, such as the bolometric temperature and the optical depth of silicates and ices.}
{We used the axisymmetric 3D radiative transfer codes RADMC and RADICAL to calculate the spectral energy distribution including silicates and ice features in a grid of models covering the range of physical properties typical of embedded and pre--main sequence sources.}
{Our set of models compares well with existing observations, supporting the assumed density parametrization and the adopted dust opacities. We show that for systems viewed at intermediate angles (25\degr\ - 70\degr ) the ``classical''  indicators of evolution are able to classify the degree of evolution of young stellar objects since they accurately trace the envelope column density, and they all agree with each other. On the other hand, edge-on system are misclassified for inclinations larger than $\sim 65$\degr $\pm 5$\degr\ ,  where the spread is mostly due to the range of mass and the flaring degree of the disk. In particular, silicate emission, typical of pre--main sequence stars with disks, turns into silicate absorption when the disk column density along the line of sight reaches $1 \cdot 10^{22}$\persc, corresponding e.g. to a $5 \cdot 10^{-3}$ \msun \ flaring disk viewed at 64\degr. A similar effect is noticed in all the other classification indicators studied: \alf , \tbol , and the \h2o\ and \co2\ ice absorption strengths. This misclassification has a large impact on the nature of the flat--spectrum sources ($\alpha \simeq 0$), whose number can be explained by simple geometrical arguments without invoking evolution. 
A reliable classification scheme using a minimal number of observations is constituted by observations of the millimeter flux with both a single dish and an interferometer.} 
{}

\keywords{stars:formation -- Stars: pre-main sequence -- ISM: dust, extinction}
\maketitle

\section{Introduction}
A correct classification of the evolutionary stages of young stellar objects (YSOs) is fundamental to understand the time--scales involved  in the star formation process.
Since protostars gradually accrete and disperse their envelope material during their lifetime, the amount of residual envelope material can be used to measure the degree of evolution of the protostellar object.
The presence of a surrounding envelope has several characteristic effects on the observables of the YSO which can then be linked to the evolutionary stage of the system. In particular, the envelope strongly affects the spectral energy distribution (SED) of the protostellar system, extinguishing and reprocessing the protostellar radiation. Thus, instead of a single blackbody emission, the SED of a young protostellar system appears as a combination of two thermal components: one corresponding to the envelope emission ($T_{\rm eff}< 100$~K) and the other that is the result of the emission of the central protostellar object ($T_{\rm eff}> 2000$~K) extinguished and scattered by the surrounding envelope. 
It is evident that measuring the ``relative intensity'' of the two components would furnish the contribution of the envelope to the whole system. 
Indeed, low--mass protostellar systems were originally classified according to the amount of extinction towards the protostar, using the ratio between near (2~\mum ) and mid-infrared fluxes (25-60~\mum ) \citep[\alf$=d{\mathrm log}(\lambda F_{\lambda})/d{\mathrm log}(\lambda)$,][]{adams1987,greene1994}, or by the relative amount of emission from the envelope, evaluated by the average frequency of emission ($\left< \nu \right>=\int \nu F_{\nu} d\nu /\int F_{\nu} d\nu$) converted to a blackbody temperature  \citep[\tbol,][]{myers1993}.
The combination of observations and theory has led to a three-class classification with class I representing the least evolved protostars with a significant fraction of the mass still left in the envelope and showing \alf$>$0  and \tbol$<$650 K;  class II, corresponding to the classical T Tauri stars with disks dominating the infrared emission and with $-2<$\alf$<0$  and 650~K$<$\tbol$<$2800~K; and class III pre-main sequence stars which show the presence of circumstellar disks only in the mid--infrared and have \alf$<-2$  and \tbol$>$2800 K. The earlier class 0 phase was added by \citet{andre1993} to include the most deeply embedded protostars that are invisible in the infrared. Finally, some authors refer to the sources with $-0.3<$\alf$<0.3$ as flat--spectrum sources and view them as transitional objects between class I and II \citep[e.g.,][]{greene1994}.

This classification has had an enormous success and a huge impact on star formation studies, but it is subject to misclassification. In particular, it does not take into account that the dissipation of the envelope is not a spherical process; in fact, while part of the material is dispersed in the outflow, the bulk of it is accreted into a disk, with the result that systems viewed edge--on would present extinction values typical of younger systems with more massive envelopes, thus leading to a misclassification.
This effect is clearly shown in the theoretical work of, e.g., \citet{whitney2003a, whitney2003b} and \citet{robitaille2006}. To emphasize the difference between the observational classification and the intrinsic properties of the YSO, \citet{robitaille2006} introduced the alternative nomenclature ``stage I, II, III", equivalent to the Lada classes, but referring to the true nature of the source regardless of its observed properties. These authors distinguish stage I from stage II  YSOs according to their accretion rate of the envelope, with the dividing line set at \mdot $_{\rm env}=10^{-6}$~\msun ~$\rm yr^{-1}$ (stage II and III are defined according to the accretion rate of the disk instead of that of the envelope). In the equations for a rotationally flattened envelope, it easily can be seen that \mdot $_{\rm env}$ is ultimately a measure of the volume density of the envelope at the centrifugal radius; for a 1~\msun \ star with a 300~AU centrifugal radius   \mdot $_{\rm env}=10^{-6}$~\msun ~$\rm yr^{-1}$ is equivalent to an H$_2$ number density of $1.8 \cdot 10^5$~\percc . \\
A possibility to observationally disentangle the extinction coming from envelopes to that produced in the disks can come from their very different temperature structures, with disks going from 30 to 1000~K and envelopes from 10 to 200~K. With such different conditions, dust grains are expected to be very different. Specifically, the envelope grains are expected to be coated by ice unlike the fraction of disk grains with temperatures above 100~K. Thus, comparing the evolutionary stage inferred by the infrared colors, like \alf , with  \tbol\, or the silicates and ice optical depths could provide a tool to distinguish between stage I YSO and edge--on disks. Including the presence of ices (\h2o\ and \co2) in our opacity tables enables us to predict not only the infrared colors, as efficiently presented in the recent literature \citep{robitaille2006}, 
but also the optical depth of spectral features, in an attempt to broaden the number of observables for YSO evolution diagnostics.

In this paper, we present model predictions for a series of YSOs covering a large range of disk mass, envelope mass and stellar luminosity, and exploring the effect of inclination on the emerging SED and spectral features.  Section~\ref{mod} presents the radiative transfer model we used and the adopted ice--coated dust properties; Section~\ref{res} compares the results of the calculation with existing data; Section~\ref{con} summarizes our conclusions.

\section{The radiative transfer model}\label{mod}
\subsection{Physical structure}

We used the three--dimensional axisymmetric radiative transfer code, RADMC \citep{dullemond2004}, to predict the continuum emission of a grid of 63 low--mass protostellar systems, each viewed at 7 different inclinations.  The code takes as input the characteristics of the star, the density structure of the surrounding material and the dust optical properties, and outputs the temperature structure and the scattering source function using a Monte Carlo technique \citep{lucy1999, bjorkman2001}. In this version of the code, scattering is supposed to be isotropic. Ray--tracing is then performed using RADICAL \citep{dullemond2000} to produce SEDs and images.
 
Given the high density reached in the center of our models the exact spectral energy distribution  of the star does not matter, because the emerging spectrum will be fully reprocessed. For this reason, we assume the (proto)--star SED to be a blackbody at 2500~K and we vary only its total luminosity in our grid between 2, 6, and 18 \lsun.\\
The density structure is the sum of three different components: {\it the disk, the envelope and the outflow cone}.
The adopted density structure for the disk has a power--law dependence along the radial coordinate and a Gaussian dependence in height, and can be expressed as 
$$ \rho_{\rm disk}(r,\theta)=\frac{\Sigma_0 \, (r/R_0)^{-1}}{\sqrt{2\pi} \, H(r)}  \exp \left\{-\frac{1}{2} \left[ \frac{r \, \cos \theta}{H(r)} \right]^2 \right\}, $$ 
where $\theta$ is the angle from the axis of symmetry.
The variation of scale-height, i.e the flaring of the disk, is described in the function $H(r)=r \cdot H_0/R_0 \cdot (r/R_0)^{2/7}$,  corresponding to the self--irradiated passive disk of \citet{chiang1997}. 
In our grid of models we fix $R_0$ to be 200~AU, $H_0$ equal to 40~AU, and we vary the total disk mass between $5 \cdot10^{-2}$, $5 \cdot10^{-3}$, and $5 \cdot10^{-4}$ \msun . \\
The envelope density follows the theoretical structure for a rotating and collapsing spheroid as derived by \citet{ulrich1976}, defined by
$$  \rho_{\rm env}(r,\theta)=\rho_0 \left(\frac{R_{\rm rot}}{r}\right)^{1.5} \left(1+\frac{\cos\theta}{\cos\theta_0}\right)^{-1/2}   \left( \frac{\cos\theta}{2 \cos\theta_0} +  \frac{R_{\rm rot}}{r}  \cos^2 \theta_0\right)^{-1}, $$ 
where $\theta_0$ is the solution of the parabolic motion of an infalling particle given by 
$r/R_{\rm rot} (\cos \theta_0 -\cos \theta) / (\cos \theta_0 \sin^2 \theta_0)=1 $, $R_{\rm rot}$ is the centrifugal radius of the envelope, and $\rho_0$ is the density in the equatorial plane at the centrifugal radius. The outer radius of the envelope is fixed to 10\,000~AU and the centrifugal radius set to 300~AU, while we vary  $\rho_0$ to cover the range of masses between 0.01 and 4.0 \msun \ in seven steps. The inner radius of the envelope is determined by the outflow cavity (see below), which crosses the equatorial plane at $\sim$20~AU.  Comparing our parametrization of the density structure to that of \citet{robitaille2006} and considering that our models have a fixed size and centrifugal radius, we can convert their assumed boundary for stage I/II of \mdot$_{\rm env}=10^{-6}$~\msun~$\rm yr^{-1}$  to a more intuitive boundary in envelope mass of \menv $=0.07$~\msun . We note  that this value would vary by factors of a few if different star mass, centrifugal radius and envelope outer radius are assumed. Given this spread, and considering that circumstellar disks more massive than one tenth of a solar mass would be gravitationally unstable, we chose to round the boundary value between stage I and II YSO to \menv $=0.1$~\msun . This boundary envelope corresponds to column densities of molecular hydrogen between 1.8 and 6$\cdot 10^{21}$~\persc\ at inclinations between 25\degr\ and 85\degr . \\
We include a streamline outflow cavity by setting the density of the region where $\cos \theta_0$ is larger than $\cos 15^\circ$ to the same value of the density of the envelope at the outer radius. This results in a funnel--shaped cavity which is conical only at large scales, where it presents a semi--aperture of 15\degr .  The total column density inside the outflow cavity can be found in Table~\ref{tabmod}.

The output temperature structure and scattering source function are then ray--traced at 7 different inclinations between 25\degr\ and 85\degr\ and spaced by 10\degr . The explored parameter space in \lstar, \mdisk , \menv , and inclination results in 441 models.

\begin{figure*}[htbp]
\centering
\resizebox{\hsize}{!}{\includegraphics{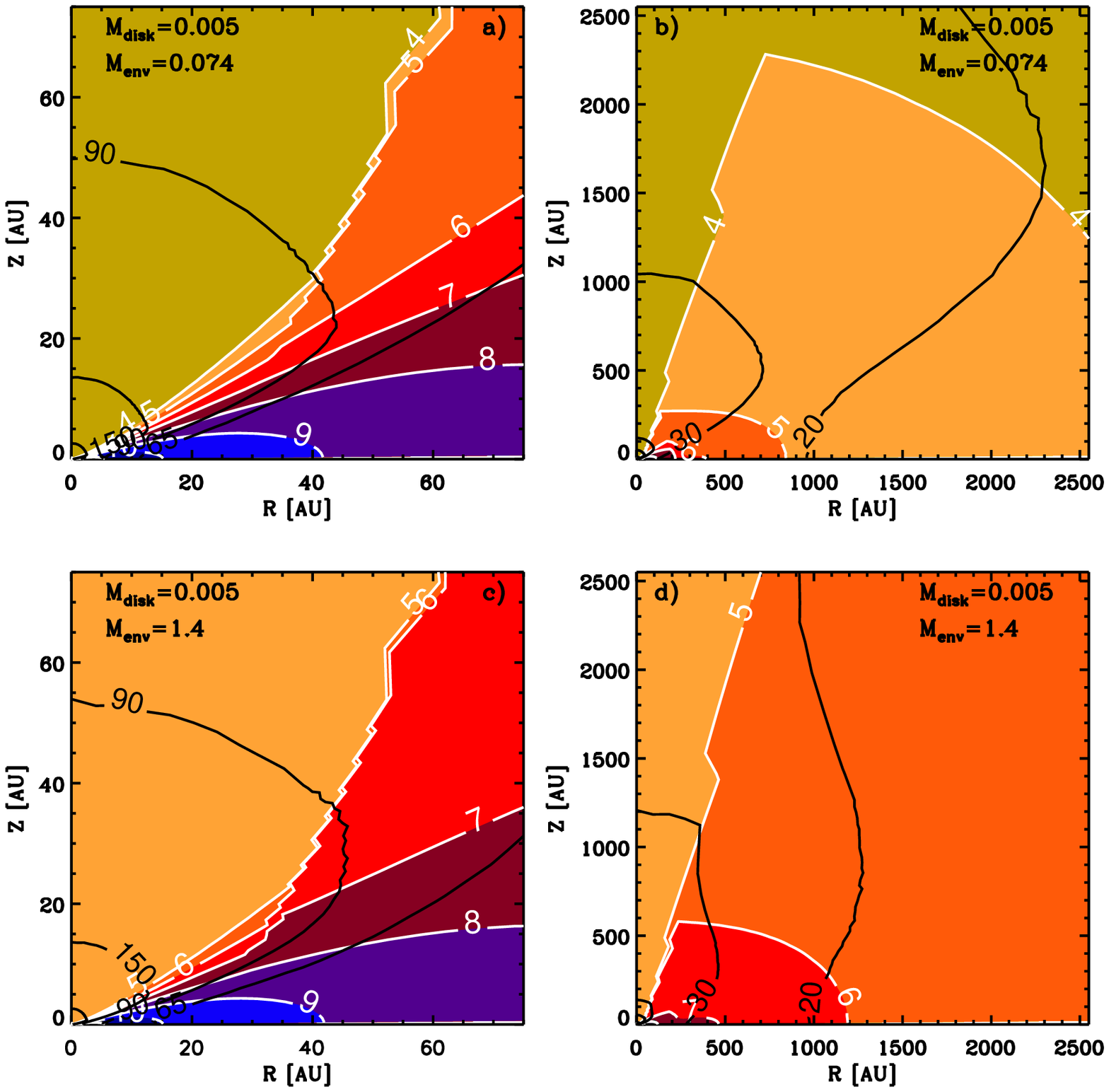}}
\caption{Density and temperature structures in two of our models. In the two top panels we show the density structure (logarithmic color-scale) and the dust temperature structure (black contours) of a model with a small envelope (\menv=0.074~\msun). The right and left panels show two different spatial scales. The two lower panels are similar to the upper two but for a model with a massive envelope (\menv=1.4~\msun). From the comparison of the two models one can see that massive envelopes are also colder than smaller ones. }
\label{Fstr}
\end{figure*}

\begin{figure*}[htbp]
\centering
\resizebox{\hsize}{!}{\includegraphics{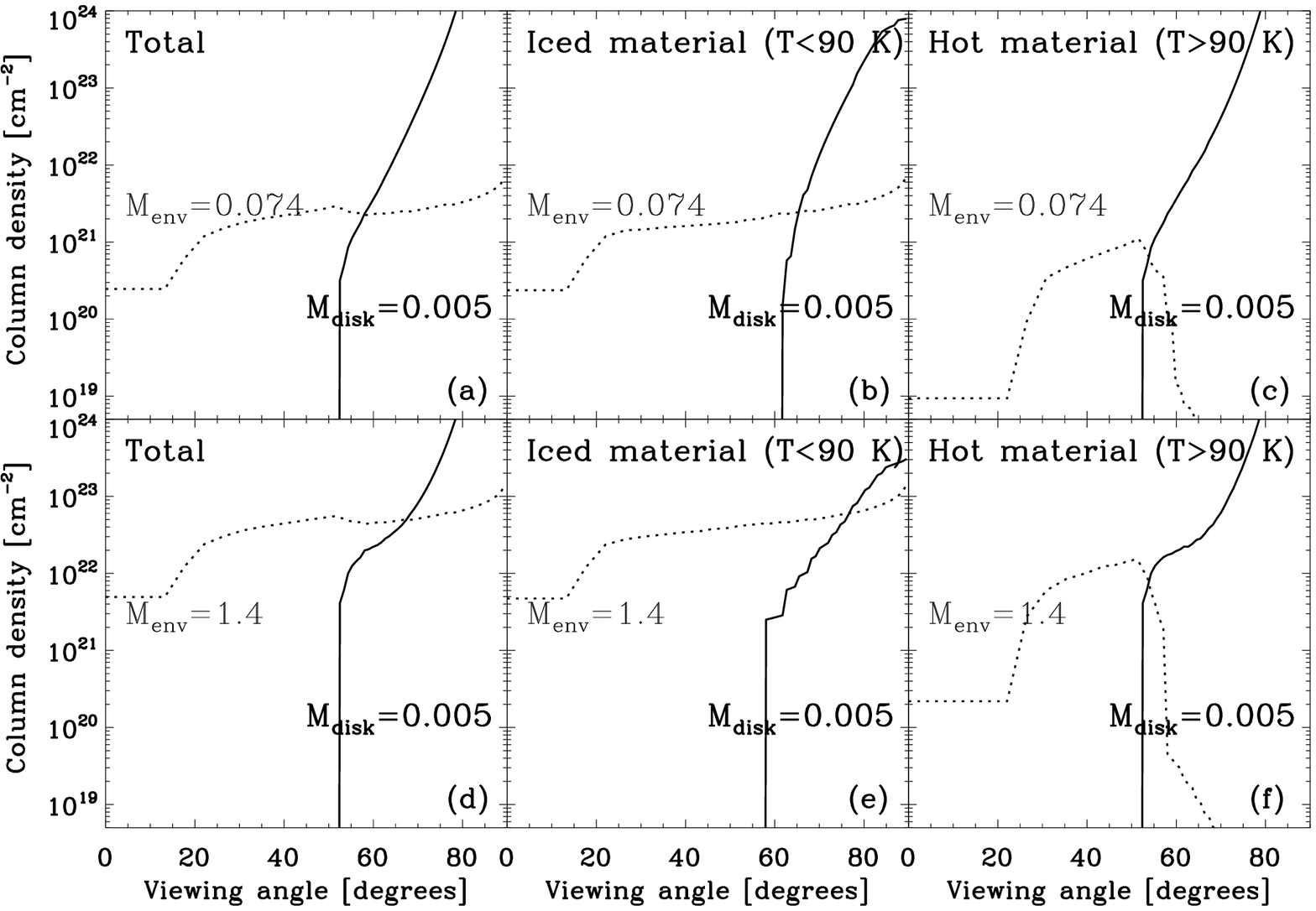}}
\caption{Column density as a function of viewing angle for two of our models, one with \menv=0.074~\msun\ (in the three top panels) and the other with \menv=1.4~\msun\ (in the three bottom panels). In all the panels, envelope material is represented with dotted line while disk material is shown with solid lines. In panels a) and d) we show the total column density, while in b) and e) we show only the column density from cells whose temperature is lower than 90~K and in c) and f) only the cells with dust temperature warmer than 90~K. Panels b) and e) thus represent the column density of material with ices, illustrating that for inclinations smaller than 65\degr --75\degr\ the envelope is the main contributor for ices along the line of sight while the disk dominates only at larger inclination.  }
\label{Fcol}
\end{figure*}

In Fig.~\ref{Fstr}, we show the temperature and density structure for two models with different envelope mass. The two structures are plotted on two different scales in the left and in the right panel to emphasize the disk and the envelope. The temperature of the most massive envelope is in general lower than that of the less massive ones. This effect is produced  by the higher density present in the inner wall of the envelope resulting in a larger fraction of energy that is absorbed closer to the star. Note also that most mass in the disk is at temperatures lower than 90~K and thus its dust will be covered by \h2o\ ice.  For the same two models, in Fig.~\ref{Fcol}, we show the contribution of the disk and of the envelope to the column density along the line of sight as a function of the viewing angle. In the left panels the total column density is presented while in the central (right) panels the contribution of iced (bare) grains is isolated. 
The disk material dominates the column density for the lines of sight that are inclined more than 60\degr --70\degr\ from the axis of symmetry. This is the case not only for the total column density but also for the column of iced grains in a pencil-beam . In fact for extreme edge--on systems the disk contribution to the column density of ices can be a few times larger than that of a massive envelope.

\subsection{The dust opacity model}\label{dustmod}
The optical properties of the dust grains used in this paper were obtained by averaging the properties of a distribution of silicates and graphite grains covered by ice mantles. 
\citet{pontoppidan2007} obtained a large grid of opacity tables for silicate and graphite grains with different size distributions and different amounts of ice mantle and compared those to the extinction observations between 1~\mum\ and 1~mm obtained by the ``Cores to Disks'' (c2d) Spitzer legacy program \citep{evans2003} and presented by  Huard et al. (in preparation).  The best fit to the extinction law of dense cores is obtained by a mixture of 71\% silicates with a size distribution similar to that of \citet{weingartner2001} with an $\alpha_{\rm WD}$-parameter\footnote{The $\alpha_{\rm WD}$ of \citet{weingartner2001} (their notation) is a parameter of the grain size distribution, not to be confused with the mid-IR slope of the SEDs} equal to -2 and turnover point for radii larger than 0.3~\mum\ covered by a mantle of ices with a water ice abundance of 3$\cdot 10^{-4}$ relative to H$_2$, and 29\% of carbonaceous grains with a size distribution similar to that of \citet{weingartner2001} characterized by a turnover point at radius of 4.5~\mum\,  an $\alpha_{\rm WD}$-parameter of -2 and a water ice abundance of 1.5$\cdot 10^{-4}$ relative to H$_2$.
This dust mixture was used in the whole grid with the exception of the regions with temperatures higher than 90~K where all the ices were removed. This was performed by running the Monte Carlo code once using only the opacities with ices in the whole grid to calculate the temperature structure, and then using this information to substitute the dust opacities in the grid cells that are found to have temperatures higher than 90~K; a new Monte Carlo run was then performed to determine the correct scattering function. Given the minimal differences in the opacities with or without ices the temperature structure is almost unaltered.
Pure \co2\ ice desorbs at a slightly different temperature than \h2o\  \citep[$\sim$65~K vs. $\sim$90~K in space, ][]{sandford1993}; since we kept the relative abundance of the two ices fixed to 20\%, the differential desorption in the temperature range 65-90~K is not taken into account in the current grid.\\

\section{Results}\label{res}
\subsection{Spectral energy distributions}\label{seds}
\begin{figure*}[htbp]
\centering
\resizebox{\hsize}{!}{\includegraphics{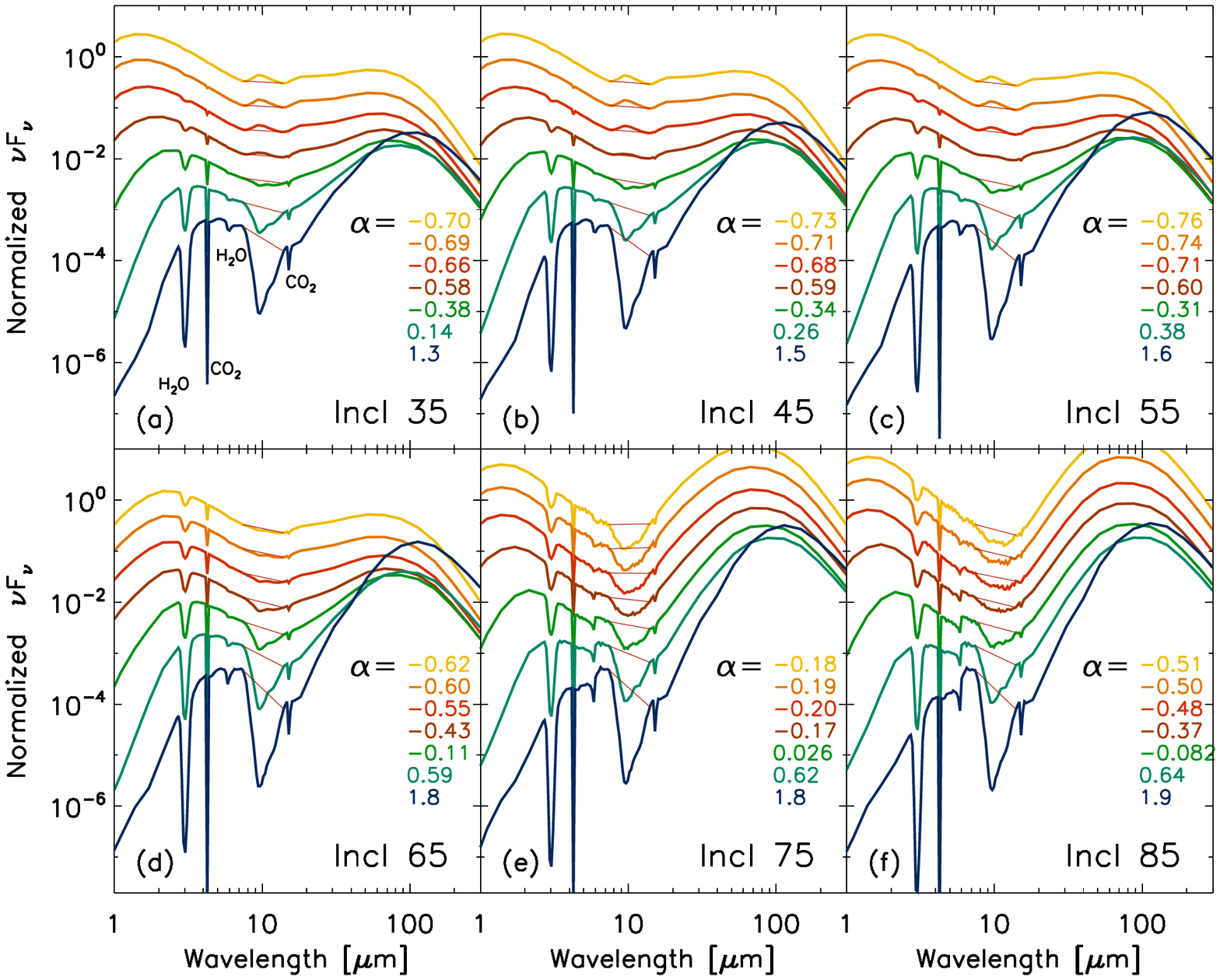}}
\caption{Normalized spectral energy distribution for a series of models with \mdisk =$5 \cdot 10^{-3}$ \msun\ and \lstar=6 \lsun . The six panels correspond to six different inclinations, and the 7 SEDs in each panel are for increasing \menv $=$0.01, 0.027, 0.074, 0.2, 0.55, 1.5 and 4~\msun\  from top to bottom. The 10~\mum \ silicate feature is in emission in the face--on objects with \menv $<$ 0.07~\msun\ while it turns in absorption for inclinations greater than $\sim 60$\degr\ (see Sect.~\ref{seds}). In the same objects \alf , the spectral index between 2.2 and 24\mum , turns from negative to positive, showing that both  indicators fail to classify edge-on disks.}
\label{Fsed}
\end{figure*}

In Fig.~\ref{Fsed}, we present the spectral energy distributions of a series of models with \mdisk =$5 \cdot 10^{-3}$~\msun\ and \lstar=6~\lsun . 
Fluxes were calculated integrating the emission inside circular apertures with radii increasing from 280~AU for near infrared wavelenghts to 10\,000~AU for millimeter wavelengths, to simulate the current observing capabilities for sources at 140~pc,  the distance of nearby star--forming regions such as the Taurus and Ophiucus molecular clouds.  
In the six panels we show six different inclinations while the sequence of SEDs shows the variation of envelope mass increasing from 0.01 to 4~\msun\  from top to bottom. 
The SEDs are normalized to be equally spaced at 8~\mum\ in order to show more clearly the differences in the silicate features. Systems closer to edge-on show a higher degree of numerical noise because the determination of the temperature and of the scattering source function in some of the cells in the mid-plane is made difficult by the high densities that limit the number of photons penetrating the innermost cells.
Systems closer to face--on, depicted in the top three panels,  show the standard sequence of SED from stage I to stage II (from bottom to top), with the silicate feature turning from absorption to emission and the ice features disappearing. The turning point from absorption to emission happens when the column density of molecular hydrogen drops below $\sim 1 \cdot 10^{22}$~\persc\ in a pencil beam, corresponding to an envelope mass of at most 0.20~\msun\ in our models. The ``evolution'' of the SED is measurable also from the change in  \alf \  (calculated from 2.2~\mum \ to 24~\mum , and reported in Fig.~\ref{Fsed}), turning from positive to negative with decreasing envelope mass, and from the bolometric temperature, increasing from 30~K to 1200~K. These results show, once again, that the Lada classification is accurate for face--on systems. More details on the models shown in Fig.~\ref{Fsed}, such as column densities, \alf , and \tbol , are reported in Table~\ref{tabmod}.
\\
In the three lower panels of Fig.~\ref{Fsed}, we start to see the effect of the disk in absorption with the disappearance of the silicate emission even in the bare disks and the corresponding negative \alf \ values. This effect is visible not only for the extreme 85\degr \ inclination but also for the more moderate value of 65\degr .  A finer grid of inclinations, performed on a model with a very small envelope (10$^{-4}$~\msun), shows that the turning point for the silicate feature
occurs at 64\degr\ for the $5 \cdot 10^{-3}$ \msun\ disk, and at 73\degr\ and 58\degr\ for the $5 \cdot 10^{-4}$ \msun\ and the $5 \cdot 10^{-2}$ \msun\ disks, respectively. This variation reflects the different inclinations where a column density of $\sim 1 \cdot 10^{22}$~\persc \ is encountered in our models. We showed above that at the same column density \alf\ also turns from positive to negative. This means that, even if we consider only the disks with the smallest masses, all the stage II sources seen in the solid angle between 70\degr\ and 90\degr\ , equal to $4\pi (cos(70)-cos(90))$ or 34\% of the total, may be mis--classified as class I or flat--spectrum sources. Recent large-scale mapping of the Perseus, Chamaleon II, and Serpens star forming regions performed by the c2d team \citep{jorgensen2006,porras2007,harvey2007} show that the ratio between flat--spectrum sources and class II objects ranges between 19\% and 47\% with an overall average of 26\%. Our simulations can explain the observed number of flat--spectrum sources  by a geometrical argument alone, without necessarily invoking evolutionary effects.

\subsection{Testing the classification schemes}
\begin{figure*}[htbp]
\centering
\resizebox{\hsize}{!}{\includegraphics{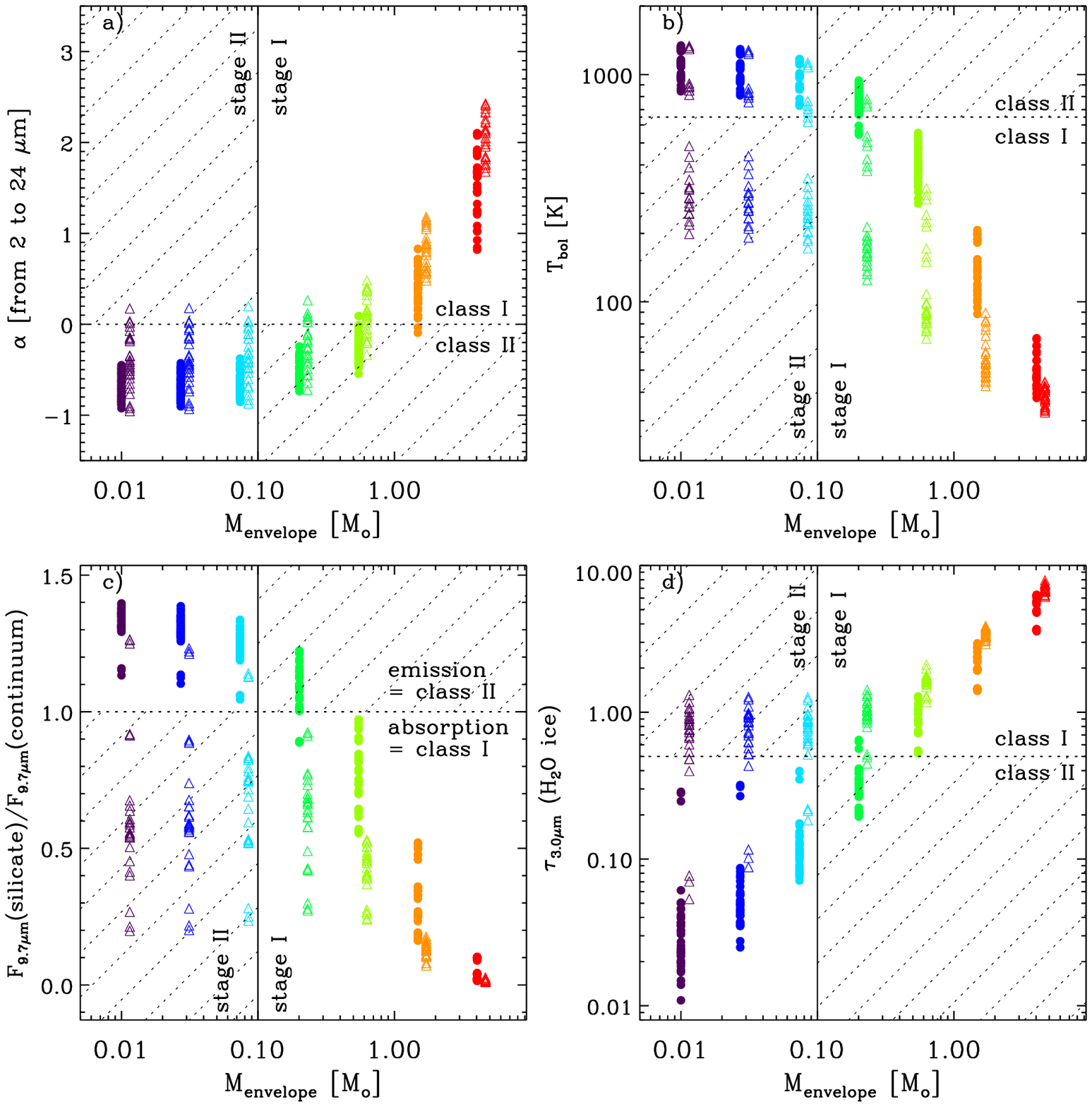}}
\caption{Traditional class indicators versus envelope mass. Colors indicate envelope mass. If the system is seen face--on (filled circles), the envelope mass is well captured by \alf , \tbol , and the depth of ices and silicates. This is not true for edge-on systems (empty triangles), and in particular for edge--on stage II YSOs which are indistinguishable from stage I objects. To visualize better the edge-on from the face-on systems in these plots, edge-on systems are offset by increasing their envelope mass by 15\%. The horizontal dotted line shows the traditional separation between class I and class II, while the vertical solid line denotes envelope masses of 0.1~\msun \ separating stage I and stage II YSOs. Sectors with a discrepancy in classification between class and stage are hatched.}
\label{Find}
\end{figure*}

To better quantify the effect of inclination in the classification of YSOs, we plot in Fig.~\ref{Find} four evolutionary indicators as a function of the mass of the envelope in our models: \alf , \tbol , the 10~\mum \ silicate feature strength, and the 3~\mum \ water ice stretch feature. The edge-on models (inclination equal to 65\degr , 75\degr \ and 85\degr ) are marked with empty triangles offset by 15\% in mass for clarity and the face-on models (inclination equal to 25\degr , 35\degr , 45\degr\ and 55\degr ) with filled circles. The colour coding separates models with different envelope mass and is thus a measure of the evolution from stage I to stage II YSOs. In all four panels, it can be seen that the face--on models form a well-defined sequence that allow us to correlate the envelope mass to the evolutionary indicators. In fact, no filled circle crosses the empirical boundary for class I/II for \menv\ $<$0.1 in any of the four panels.
This is  not the case for the edge-on models, which appear less sensitive to the decrease of envelope mass, resulting in similar values of \alf , \tbol , and silicate and ice features for both stage I and stage II objects. As a result, edge-on disks can be erroneously classified as class I YSOs when using these evolutionary indicators. This is because \alf , \tbol \ and the silicate feature are sensitive to the column density along the line of sight, and the additional column of material present in the disk has a similar effect on the SED  as a more massive envelope. 
However, even though the column density through an edge--on disk can exceed the column density of the most massive envelope by one or two orders of magnitude (see Table~\ref{tabmod}), none of the edge--on disks (\menv $<$0.1) present values of \alf\ as high as those observed in the models with \menv = 4~\msun . This is due to the fact that a large part of the 2~\mum\ flux is scattered back into the line of sight through the material above the disk and this scattered light is never absorbed again, leading to large 2~\mum\ flux and consequently  much smaller \alf\ values than those measured in massive envelopes, where the extinction at 2~\mum\ comes  from the sum of absorption and scattering. In other words scattering allows lines of sight with low absorption at 2~\mum\ for edge--on disks unavailable for massive envelopes.
\\
The effect of inclination on \tbol\  (Fig.~\ref{Find}b) grows in strength from stage I to stage II YSO.
We note that even for the least massive disk that we studied,  at large inclinations \tbol\ crosses the limit for class~I sources  (650~K) and would lead to a classification as class I. Given the much larger number of observed class II than class I sources in star-forming regions \citep[392 vs 104 in the combined samples of ][]{jorgensen2006, porras2007, harvey2007}, a small incidence  of YSOs with protostellar disks misclassified as class I when seen edge-on could dramatically change the timescale of the embedded phase versus the protostellar disk phase. Taking just $\sim$10\% of stage II objects that present values of \alf\ typical of class I when seen edge-on, simple algebra shows that a measured ratio of $n$(class I)$/n$(class II) $=$ 26\% would yield an intrinsic ratio of $n$(stage I)$/n$(stage II) $=$ [$n$(class I)$-n$(edge-on disk)]$/$[$n$(class II)$+n$(edge-on disk)] $=$ 13\%, half of the measured value. 
\\
The effect of scattering is less strong in the silicate feature vs. the envelope mass (Fig.~\ref{Find}c) than in the \alf\ panel. This is because silicates probe the entire column density, in contrast with ices which evaporate in the high density and high temperature inner part of the disk. Still, more massive envelopes produce larger values of silicate absorption than edge--on disks. This is not a column density effect but is due to the fainter overall continuum which in edge--on disks is increased by scattering. 
\\
For the optical depth of the 3~\mum\ \h2o\ ice feature (Fig.~\ref{Find}d) we also note that edge-on disks present values much lower than massive envelopes despite higher column densities (see Fig.~\ref{Fcol}), again an effect of evaporation of ices combined with scattering of the surrounding continuum.
A plot for the 15.2~\mum\ feature of \co2\ is qualitatively similar to that for \h2o\ with the exception that edge--on disks now reach values typical of 1.5~\msun\ envelopes, due to the smaller effect of scattering in the mid--IR. For the \h2o\ ices there is no ``traditional'' boundary between class I and II. We adopted $\tau=0.5$ because in this way the boundary crosses the \menv\ = 0.2~\msun\ models as it does for all the other indicators. In a similar fashion, below we chose the boundary for the \co2\ ice feature to be at $\tau=0.1$.

\begin{figure*}[htbp]
\centering
\resizebox{!}{22cm}{\includegraphics{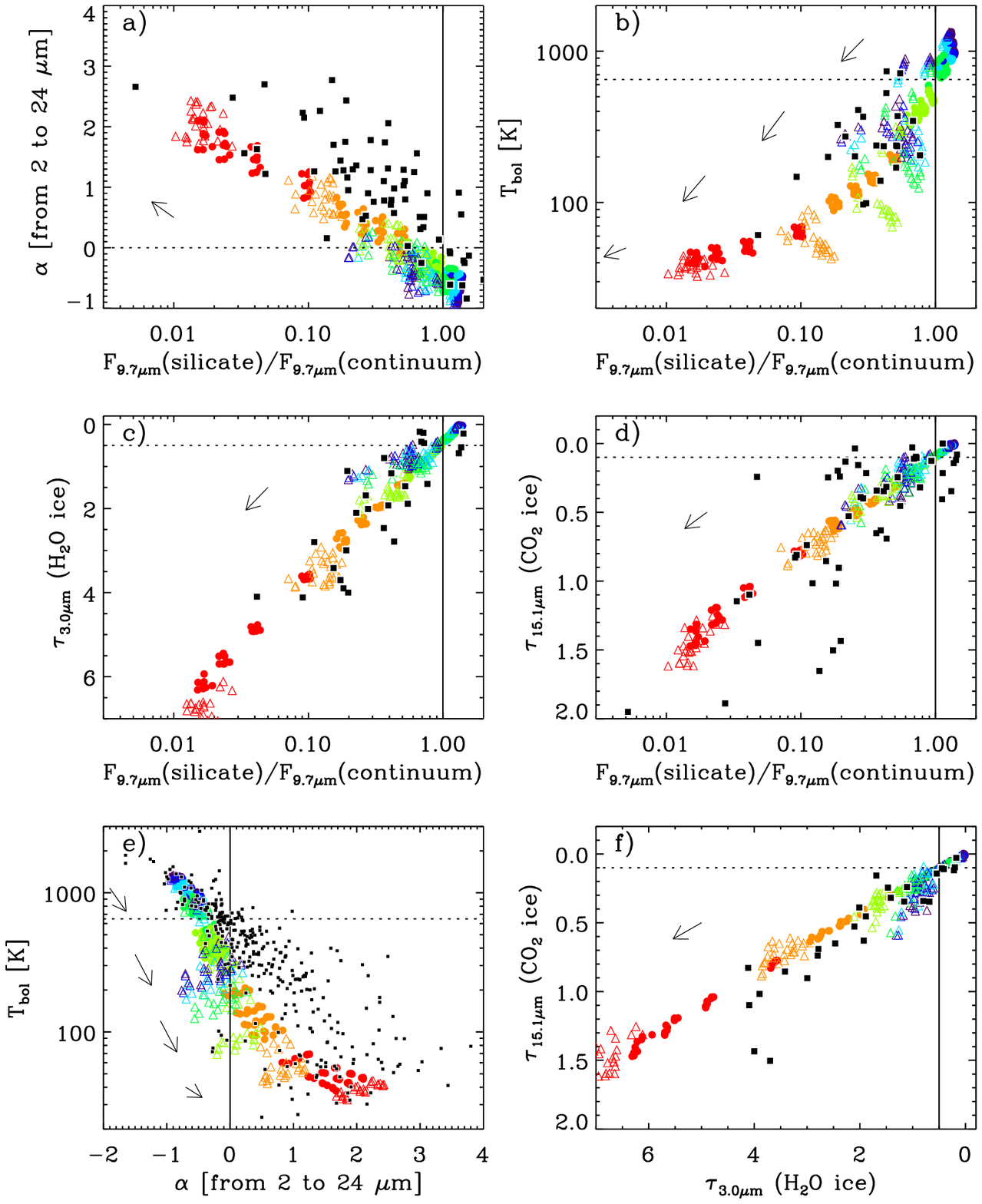}}
\caption{Cross-correlation of various indicators. In panels a to d we compare the silicate feature depth with \alf ,\tbol , and the optical depth of water-ice and \co2\ ice, while in panel e \alf \ is plotted versus \tbol \ and in panel f the two ices optical depths are cross correlated. In each panel observations from the literature are plotted as black squares. The boundary between class I and class II objects is shown with a vertical solid line and a horizontal dotted line. Face-on (filled circles) and some edge-on systems (empty triangles) form a well defined sequence that allow to relate the different evolutionary indicators. Color coding is the same as in Fig.~\ref{Find} and indicates envelope mass. Besides the correlations in panels a to d, a slightly different trend seems to be found for edge-on disks, which present the heaviest geometrical effect. The effect on the evolutionary indicators due to a $10^{22}$~\persc \ foreground layer are represented with arrows in each panel. Since the effect of the foreground on \tbol\ is not uniform we show the reddening vectors in 4 different ranges of \tbol\ (panels b and e) . }
\label{Ficei}
\end{figure*}

From the analysis of the results in Fig.~\ref{Find}, it is evident that geometry has a different effect on the four indicators, so cross--correlating two evolutionary indicators could reveal different trends for the face--on and for the edge--on objects.
This is shown in Fig.~\ref{Ficei}, where the silicate feature strength is plotted against \alf , \tbol , the optical depths of water ice and \co2\ ice in panels a--d, while \alf \ versus \tbol \ is shown in panel e and the optical depths of the ices of water and \co2\ are cross--correlated in panel f. 
The color coding and the shape coding are the same as in Fig.\ref{Find}. In each plot we added as small, black squares observations from the literature, in particular for sources studied by the c2d team with the Spitzer IRS. Values and references can be found in Table~\ref{tabobs} in the online material. The models seem to reproduce fairly well the observational trends over the entire range of variables.
Panels a-d show that indeed edge-on disks (open triangles) may not follow the same trends as the other models. If compared with embedded or face-on objects they systematically show deeper silicate absorption than  would have been expected from the other indicators. As shown above this is due to the depth of the silicate absorption feature being more sensitive to the total column density and less affected by scattering. This effect unfortunately is quite small and smaller than the current observational error on the indicators. The strong effect of scattering on the ice and silicate features makes it impossible to select the edge-on disks from the embedded YSOs from the different temperature structure along the line of sight. \\
The \tbol\--\alf\ plane (Fig.~\ref{Ficei}e) is the one that is most populated by data. Most measurements come from the systematic work of Enoch et al. (in prep.) which makes use of  2MASS, Spitzer and Bolocam 1.1~mm data to calculate \tbol\ and \alf\ for all the YSOs in the Perseus, Serpens and Ophiucus star-forming regions studied by c2d. The scatter in the observations is larger than in the models, which could be the result of foreground layers, incomplete sampling of the SED, source confusion or the fact that in case of non-detections at 2.2~\mum\ Spitzer bands were used to calculate \alf . Still, the general trend is reproduced reasonably well. Applying the effect of a foreground cloud (see arrows in Fig.\ref{Ficei} and Sect.~\ref{disc}) shows that while the sources in the bottom left corner can be still explained by these models, the sources that lie above the trend (1$<$\alf $<$3 and 100~K$<$\tbol $<$600~K ) cannot and need further explanation.

\subsection{Sub--millimeter indicators}
\begin{figure*}[htbp]
\centering
\resizebox{\hsize}{!}{\includegraphics{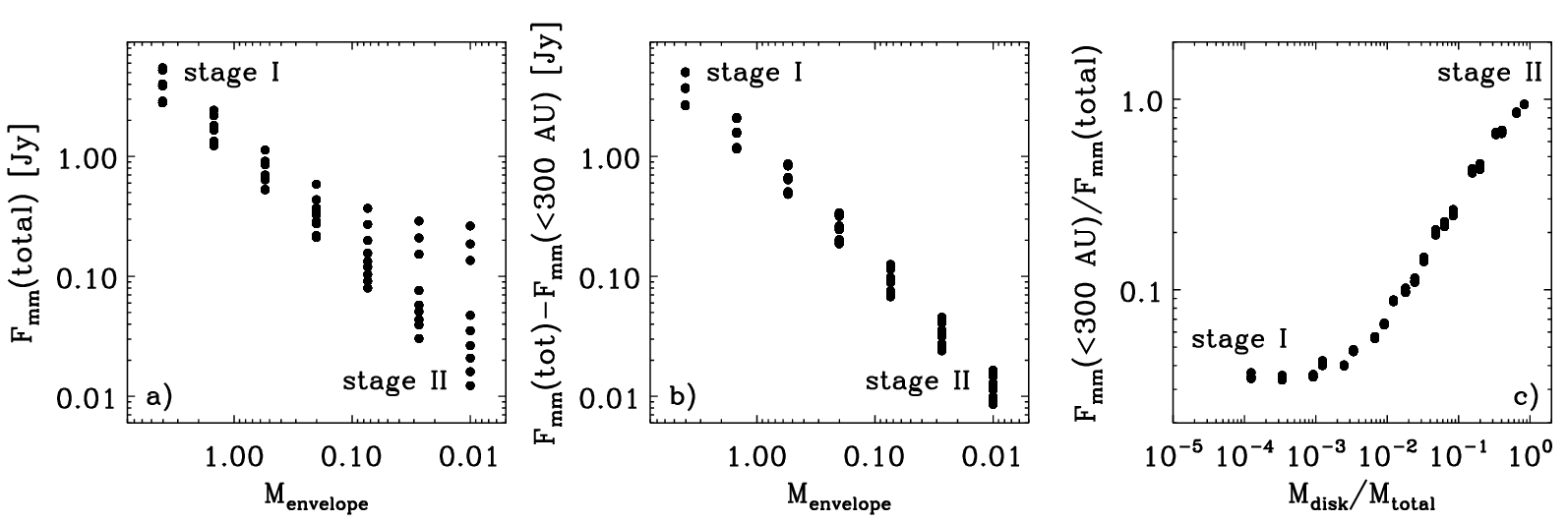}}
\caption{Millimeter flux versus envelope mass. In panel a the envelope mass is seen to be correlated to the total flux at 1.2~mm (calculated assuming a distance of 140~pc). The envelope mass can be measured through this optically thin emission, but this can hardly be used to disentangle the edge-on disks from the low-mass envelopes, because in the transition between stage I and stage II, the masses of the disk and of the envelope become comparable, as are the two emissions (as shown in the right part of the plot). In panel b, we correct for the flux coming from the inner 300~AU ($\sim$2\arcsec \ at 140~pc)  (obtainable with mm-interferometry) showing a much tighter correlation. In panel c we substitute the envelope mass with the ratio between disk and total mass and show that this correlates perfectly with the ratios between mm--interferometry and single--dish fluxes.}
\label{Fm}
\end{figure*}

An alternative to directly measure the envelope mass is through observation of optically thin tracers like the continuum at 1.2~mm. In  Fig.~\ref{Fm}a, we show the predictions for the total 1.2~mm emission versus the mass of the envelope in our grid of models assuming an arbitrary distance of 140~pc. A correlation is seen when the envelope is more massive than 0.1~\msun , but the dependence of the millimeter flux on the mass of the envelope becomes flat for stage II YSOs. 
This occurs because the disk starts to emit comparable amounts of flux. The single--dish flux can be corrected for the contribution from the disk by subtracting the flux measured in a much smaller region (Fig.~\ref{Fm}b),  such as the one obtained by (sub)mm-interferometry. The ratio of single--dish over interferometer flux is perfectly correlated with envelope over disk mass (Fig.~\ref{Fm}c), and offers a sensitive probe of evolutionary stage independent of orientation. Using the visibility amplitude  versus projected baseline length can further refine the distinction between envelope and disk. With the increase in sensitivity of (sub)mm-interferometry currently reached at Plateau de Bure, SMA and CARMA, we expect these observations to be routinely done to check the true nature of class I objects.

\subsection{Discussion}\label{disc}
In our grid, we fixed some parameters that affect the density distribution and so may affect the results.
These were: the envelope centrifugal radius (fixed here at 300~AU), the disk and the envelope outer radii (fixed at 200~AU and 10\,000~AU), the scale height distribution of the disk (fixed here to be 20\% at 200~AU and decreasing with radius to the power of 2/7). Their effects are degenerate with those of changing the disk and the envelope masses, and for this reason we chose to keep them fixed. Here we explain  how they change the results. \\
Decreasing the centrifugal radius  results in  more peaked and more spherical envelopes, so for the same envelope mass  we would get a larger column density. Varying the centrifugal radius by a factor of 2 changes the column density by 30\% in the same direction.
Varying the disk and the envelope outer radii while keeping the same mass decreases the local volume density and thus decreases the column density along the line of sight. Doubling the outer radius reduces the column density by a factor of $\sim$3, and vice-versa.  
Finally, varying the scale height distribution may change the boundary inclination where, e.g., silicates turn from emission to absorption at a given mass. This boundary value was shown to range between 60\degr\ and 70\degr\ depending on the disk mass. 
An increase of 10\%\  in the scale height reduces the boundary value by 5\degr ; larger changes in the disk scale height are unlikely because of stability considerations.

We can also calculate the effect of a foreground layer of a given column density, assumed to be dense enough to have ices throughout, by  extinguishing the calculated SED and comparing with the original SED. As an example, having a foreground layer of $10^{22}$~\persc \ ($\mathrm A_V \sim 5.5$) would result in lowering \alf\ by 0.21, dividing the silicate feature to continuum ratio by 1.46, and increasing the \h2o\ and \co2\ optical depths by 0.5 and 0.1 respectively.  
In the case of \tbol\ the effect of a foreground layer is neither linear nor logarithmic and depends on the details of the SED; on average, a \tbol\ of 1200~K would decrease by 350~K after being extinguished by a foreground layer of $10^{22}$~\persc \, while the respective reductions for a \tbol\ of 600~K and 200~K are $\sim$200~K and $\sim$75~K. These reddening values relative to a $10^{22}$~\persc \ foreground layer are represented with arrows in Fig.~\ref{Ficei}.

Because of the rough parametrization of the density in the outflow cavity, we do not present any model results for face-on disks with $ i<$25\degr . Test calculations show that widening of the outflow cavity to 25\degr\ has little impact on the observables for $i>$30\degr , but strongly affects them at $i$=0-20\degr . Consistent with previous work \citep{calvet1994,whitney2003a} we find that Class I objects, seen below 15\degr , show flat spectra and silicates in emission, even for envelope masses up to 1~\msun .

We did not explore different (proto) stellar temperatures. For this reason, our models cannot be representative of high-mass YSOs with stellar temperatures higher than 5000~K, particularly in the case of the less embedded objects where the central radiation is not fully reprocessed \citep[for a better description of this effect see, e.g.,][]{whitney2004}.

\section{Conclusions}\label{con}
We used a Monte Carlo-based radiative transfer code to calculate the emission for a grid of models representing stage I and II young stellar objects: embedded YSOs and pre--main sequence stars with disks. The aim was to study the behaviour of  classification parameters that are used to distinguish YSOs embedded in envelopes from those surrounded by disks only.
We can summarize our results in these main points:
\begin{itemize}
\item The classical indicators of YSO evolution, \alf ,\tbol , and the silicate emission/absorption can describe very well the disappearance of the protostellar envelope for those systems seen at intermediate inclinations (approximately from 25\degr\ to 65\degr ), accounting for more than half of the sources. Interestingly the transition between class I and class II measured with \alf , \tbol , or with the silicate feature happens for the same line of sight column density of $\sim1 \cdot 10^{22}$~\persc \  (corresponding to \menv $\sim$ 0.2~\msun ), which means that these methods are well calibrated relative to each other.
\item  Each of these indicators confuses pre--main sequence stars with edge-on disks with heavily embedded YSOs if the column density exceeds $1 \cdot 10^{22}$~\persc ; this column density is reached for inclinations greater than 60\degr --70\degr , depending on the mass of the disk.
\item Even a pre--main sequence star surrounded by disk of only $5 \cdot 10^{-4}$ \msun\ is classified as a flat--spectrum source for inclinations greater than 70\degr. This means that 34\% of the stage II sources are misclassified and could account for the entire population of flat--spectrum sources observed in nearby star--forming regions. If 10\% of pre--main sequence disk objects are misclassified as class I, the derived time-scale for the embedded phase is overestimated by a factor of two, when using the ratio of class I/II blindly.
\item \co2\ and \h2o\ ice optical depth correlate well with \alf\ and  \tbol, providing an alternative way to classify YSOs. For edge-on sources, the strong effect of scattering saturates the continuum around the features, thus affecting the feature strength and hiding the effect of temperature structure along the line of sight.
\item A combination of  single dish and interferometer millimeter continuum observations is able to trace accurately, and with the minimum number of observations, the amount of remaining envelope in emission and thus the true evolutionary stage of the YSO.
\end{itemize}

\begin{acknowledgements}

A.C. was supported by a fellowship from the European Research Training Network
"The Origin of Planetary Systems'' (PLANETS, contract number HPRN-CT-2002-00308)
at Leiden Observatory and by a Marie Curie Intra-European Fellowship from the European Community (contract number FP6-024227) at Observatorio Astron\'omico Nacional. M.R.H. and A.C. are supported by a VIDI grant from the Netherlands Organization for Scientific Research. KMP is supported by NASA through Hubble Fellowship grant \#01201.01 awarded by the Space Telescope Science Institute, which is operated by the Association of Universities for Research in Astronomy, Inc., for NASA, under contract NAS 5-26555.
We are grateful to Melissa Enoch for providing her \alf \ and \tbol\ measurements before publication, and to  Neal Evans, Fred Lahuis, Jes J{\o}rgensen, Jean-Charles Augereau and the referee, Barbara Whitney, for enlightening discussions.

\end{acknowledgements}

\begin{table*}
\caption{Observables for the models presented in Fig.~\ref{Fsed}} 
\label{tabmod}

\begin{tabular}{lccccccccc}
\hline\hline 
			 				 &				    &$i=5$\degr\ & $i=25$\degr \ &  $i=35$\degr \ & $i=45$\degr & $i=55$\degr \ &  $i=65$\degr \ & $i=75$\degr \ & $i=85$\degr \  \\ 
\hline 
$M_{\rm disk}=5\cdot10^{-3}$ & $N$(H$_2$) in disk [\persc ] 	&  0       & 0  	  & 0		 & 3.61(16) & 5.56(20) & 1.07(22) & 1.76(23) & 2.75(25) \\
$M_{\rm env }= 0.010$		 & $N$(H$_2$) in envelope [\persc ] & 3.17(19) & 1.78(20) & 2.33(20) & 2.95(20) & 2.60(20) & 2.69(20) & 3.40(20) & 5.03(20) \\
 							 & $\alpha$		 					&		   & $-$0.68  & $-$0.70  & $-$0.73  & $-$0.76  & $-$0.62  & $-$0.18  & $-$0.51  \\  
 							 & \tbol \ [K]			    		&		   &	1092  &    1107  &    1129  &	 1142  &	 879  & 	311  &     287  \\  
$M_{\rm env }= 0.027$		 & $N$(H$_2$) in envelope [\persc ] & 8.62(19) & 4.84(20) & 6.34(20) & 7.78(20) & 8.08(20) & 7.99(20) & 9.55(20) & 1.44(21) \\
 							 & $\alpha$			    			&		   & $-$0.67  & $-$0.69  & $-$0.71  & $-$0.74  & $-$0.60  & $-$0.19  & $-$0.50  \\  
 							 & \tbol \ [K]			    		&		   &	1062  &    1071  &    1090  &	 1098  &	 836  & 	293  &     272  \\  	   
$M_{\rm env }= 0.074$		 & $N$(H$_2$) in envelope [\persc ] & 2.34(20) & 1.31(21) & 1.72(21) & 2.09(21) & 2.49(21) & 2.22(21) & 2.68(21) & 4.09(21) \\
 							 & $\alpha$			    			&		   & $-$0.65  & $-$0.66  & $-$0.68  & $-$0.71  & $-$0.55  & $-$0.20  & $-$0.48  \\  
 							 & \tbol \ [K]			    		&		   &	 986  & 	981  &     988  &	  984  &	 729  & 	251  &     233  \\  
$M_{\rm env }= 0.20$		 & $N$(H$_2$) in envelope [\persc ] & 6.37(20) & 3.57(21) & 4.68(21) & 5.66(21) & 7.19(21) & 6.17(21) & 7.48(21) & 1.15(22) \\
 							 & $\alpha$			    			&		   & $-$0.59  & $-$0.58  & $-$0.59  & $-$0.60  & $-$0.43  & $-$0.17  & $-$0.37  \\  
 							 & \tbol \ [K]			    		&		   &	 804  & 	767  &     749  &	  720  &	 504  & 	172  &     159  \\  
$M_{\rm env }= 0.55$		 & $N$(H$_2$) in envelope [\persc ] & 1.73(21) & 9.72(21) & 1.27(22) & 1.54(22) & 1.87(22) & 1.77(22) & 2.09(22) & 3.23(22) \\
 							 & $\alpha$			    			&		   & $-$0.43  & $-$0.38  & $-$0.34  & $-$0.31  & $-$0.11  &   0.026  &$-$0.082  \\  
 							 & \tbol \ [K]			    		&		   &	 486  & 	414  &     374  &	  332  &	 224  & 	 89  &  	82  \\  
$M_{\rm env }= 1.5$			 & $N$(H$_2$) in envelope [\persc ] & 4.71(21) & 2.64(22) & 3.46(22) & 4.17(22) & 5.01(22) & 7.02(22) & 5.81(22) & 8.50(22) \\
 							 & $\alpha$			    			&		   &  -0.038  &    0.14  &    0.26  &	 0.38  &	0.59  &    0.62  &    0.64  \\  
 							 & \tbol \ [K]			    		&		   &	 188  & 	139  &     117  &	   99  &	  75  & 	 51  &  	48  \\  
$M_{\rm env }= 4.0$			 & $N$(H$_2$) in envelope [\persc ] & 1.28(22) & 7.18(22) & 9.40(22) & 1.13(23) & 1.35(23) & 1.74(23) & 1.74(23) & 2.38(23) \\
		        			 & $\alpha$    		    	    	&		   &	0.85  & 	1.3  &     1.5  &	  1.6  &	 1.8  & 	1.8  &     1.9  \\  
		        			 & \tbol \ [K]     	    	    	&		   &	  63  & 	 51  &  	46  &	   42  &	  40  & 	 38  &  	37  \\  

\hline 
\end{tabular}

\end{table*}

\onltab{2}{
\begin{table*}
\caption{Observed indicators for various low--mass YSOs from the literature for sources in the c2d sample. } 
\label{tabobs}

\begin{tabular}{lcccccc}
\hline\hline 
Source & $F_{9.7}/F_{continuum}$ & \alf (2.2-24\mum ) & \tbol & $\tau$\h2o & $\tau$\co2 & Ref.\\ 
\hline 
 IRAS 04169+2702 &   0.51 &	1.29 &     170 &  ...	 &  ...    & 1,3 \\
IRAS 04181+2654A &   0.68 &	0.31 &     346 &  ...	 &  ...    & 1,3 \\
 IRAS 04239+2436 &   0.52 &	1.10 &     236 &  ...	 &  ...    & 1,3 \\
      Haro 6-10A &   0.41 &	0.71 &    ...  &  ...	 &  ...    & 1 \\
      Haro 6-10B &   0.64 &	1.56 &    ...  &  ...	 &  ...    & 1 \\
 IRAS 04287+1801 &   0.29 &	1.78 &      97 &  ...	 &  ...    & 1,3 \\
        Lk Ca 15 &   3.18 &    -0.90 &    ...  &  ...	 &  ...    & 1 \\
 IRAS 04381+2540 &   0.39 &	1.31 &     139 &  ...	 &  ...    & 1,3 \\
     Hen 3-600 A &   1.39 &    -0.62 &    ...  &  ...	&  ...    & 1 \\
 IRAS 14050-4109 &   2.85 &    -0.68 &    ...  &  ...	&  ...    & 1 \\
       HD 163296 &   2.33 &    -0.58 &    ...  &  ...	&  ...    & 1 \\
       HD 179218 &   1.56 &    -0.13 &    ...  &  ...	&  ...    & 1 \\
          WW Vul &   2.65 &    -0.57 &    ...  &  ...	&  ...    & 1 \\
       HD 184761 &   2.45 &    -1.38 &    ...  &  ...	&  ...    & 1 \\
      L1448 IRS1 &   0.72 &    0.34  &    ...  &   0.2  &   0.064 & 2 \\
 IRAS 03235+3004 &   0.17 &    1.44  &    ...  &   3.7  &   1.50 & 2 \\
 IRAS 03245+3002 &   0.047 &    2.70 &    ...  &  ...   &   0.24 & 2 \\
      L1455 SMM1 &   0.19 &    2.41  &    ...  &   3.0  &   0.90 & 2 \\
          RNO 15 &   1.42 &    -0.21 &    ...  &   0.22 &   0.11 & 2 \\
      L1455 IRS3 &   0.93 &    0.98  &    ...  &  ...   &   0.12 & 2 \\
 IRAS 03254+3050 &   0.28 &    0.90  &    ...  &   2.0  &   0.39 & 2 \\
 IRAS 03271+3013 &   0.39 &    2.06  &    ...  &   1.9  &   0.63 & 2 \\
 IRAS 03301+3111 &   0.67 &    0.51  &    ...  &   0.17 &   0.028 & 2 \\
            B1-a &   0.033 &   1.87  &    ...  &  ...   &   1.15 & 2 \\
            B1-c &   0.005 &   2.66  &    ...  &  ...   &   1.95 & 2 \\
            B1-b &   0.14 &    0.68  &    ...  &  ...   &   1.65 & 2 \\
 IRAS 03439+3233 &   0.72 &    0.51  &    ...  &   0.45 &   0.10 & 2 \\
 IRAS 03445+3242 &   0.36 &    0.78  &    ...  &   0.80 &   0.34 & 2 \\
       L1489 IRS &   0.36 &    1.10  &    238  &   2.5  &   0.65 & 2,3 \\
IRAS 04108+2803B &   0.77 &    0.90  &    205  &   1.4  &   0.32 & 2,3 \\
 IRAS 04239+2436 &   0.41 &    0.79  &    236  &   1.2  &   0.35 & 2,3 \\
        DG Tau B &   0.19 &    1.16  &    ...  &   1.1  &   0.24 & 2 \\
        HH46 IRS &   0.15 &    1.70  &    ...  &   3.4  &   0.85 & 2 \\
 IRAS 12553-7651 &   0.31 &    0.76  &     99  &  ...   &   0.21 & 2,4 \\
 IRAS 13546-3941 &   1.12 &    -0.06 &    ...  &  ...   &   0.40 & 2 \\
 IRAS 15398-3359 &   0.048 &    1.22  &    61  &  ...   &   1.45 & 2,4 \\
        Elias 29 &   0.27 &    0.53  &    409  &   1.7  &   0.16 & 2,3 \\
CRBR 2422.8-3423 &   0.23 &    1.36  &    ...  &   2.1  &   0.53 & 2 \\
          RNO 91 &   0.55 &    0.03  &    715  &   1.9  &   0.45 & 2,3 \\
 IRAS 17081-2721 &   1.37 &    0.55  &    ...  &   0.55 &   0.14 & 2 \\
        B59 YSO5 &   0.12 &    2.26  &    ...  &  ...   &   1.02 & 2 \\
  2MASSJ17112317 &   0.027 &   2.48  &    ...  &  ...   &   1.89 & 2 \\
           EC 74 &   0.69 &    -0.25 &    ...  &   0.41 &   0.11 & 2 \\
           EC 82 &   3.06 &    0.38  &    ...  &  ...   &   0.09 & 2 \\
         SVS 4-5 &   0.11 &    1.26  &    ...  &   2.8  &   0.74 & 2 \\
           EC 90 &   1.13 &    -0.09 &    ...  &  ...   &   0.21 & 2 \\
           EC 92 &   1.31 &    0.91  &    ...  &   0.68 &   0.35 & 2 \\
             CK4 &   1.45 &    -0.25 &    ...  &  ...   &   0.08 & 2 \\
     R CrA IRS 5 &   0.43 &    0.98  &    736  &   2.8  &   0.69 & 2,4 \\
        HH100 IRS&   0.52 &    0.80  &    373  &   1.5  &   0.24 & 2,4 \\
     CrA IRS 7 A &  0.091 &    2.23  &    ...  &   4.1  &   0.83 & 2 \\
     CrA IRS 7 B &  0.041 &    1.63  &    ...  &   4.1  &   1.10 & 2 \\
      CrA IRAS32 &  0.093 &    2.15  &    148  &  ...   &   0.81 & 2,4 \\
       L1014 IRS &   0.18 &    1.28  &    ...  &   3.9  &   1.02 & 2 \\
 IRAS 23238+7401 &   0.20 &    0.95  &    ...  &   4.0  &   1.43 & 2 \\
           WL 12 &   0.19 &     1.75 &     325 &  ...   &   0.19 & 5,3 \\
            WL 6 &   0.29 &     1.28 &     369 &  ...   &   0.40 & 5,3 \\
          IRS 63 &   0.43 &     0.57 &     529 &  ...   &   0.31 & 5,3 \\
 IRAS 04302+2247 &   0.25 &     0.47 &     202 &  ...   &   0.036& 5,3 \\
     GSS 30 IRS1 &   0.21 &     1.30 &     273 &  ...   &   0.13 & 5,3 \\
       L1535 IRS &   0.16 &     1.26 &     200 &  ...   &   0.25 & 5,3 \\
     2MASS 16282 &   1.13 &    -0.61 &    ...  &  ...   &   0.00 & 5 \\

\hline 
\multicolumn{7}{l}{ (1) \citet{kessler2005}  } \\
\multicolumn{7}{l}{ (2) \citet{boogert2007} } \\
\multicolumn{7}{l}{ (3) \citet{chen1995} } \\
\multicolumn{7}{l}{ (4) \citet{chen1997} } \\
\multicolumn{7}{l}{ (5) the c2d legacy data archive } \\
\end{tabular}

\end{table*}
}


\begin{thebibliography}{}

\bibitem[Adams et al.(1987)]{adams1987} Adams, F.~C., Lada, 
C.~J., \& Shu, F.~H.\ 1987, \apj, 312, 788 

\bibitem[Andr{\'e} et al.(1993)]{andre1993} Andr{\'e}, P., Ward-Thompson, 
D., \& Barsony, M.\ 1993, \apj, 406, 122 

\bibitem[Bjorkman \& Wood(2001)]{bjorkman2001} Bjorkman, J.~E., \& Wood, K.\ 2001, \apj, 554, 615 

\bibitem[Boogert et al.(in prep.)]{boogert2007} Boogert, A.~C.~A.~B., et al. 2007, in prep.

\bibitem[Calvet et al.(1994)]{calvet1994} Calvet, N., Hartmann, L., Kenyon, S.~J., \& Whitney, B.~A.\ 1994, \apj, 434, 330 

\bibitem[Chen et al.(1995)]{chen1995} Chen, H., Myers, P.~C., 
Ladd, E.~F., \& Wood, D.~O.~S.\ 1995, \apj, 445, 377 

\bibitem[Chen et al.(1997)]{chen1997} Chen, H., Grenfell, T.~G., 
Myers, P.~C., \& Hughes, J.~D.\ 1997, \apj, 478, 295 

\bibitem[Chiang \& Goldreich(1997)]{chiang1997} Chiang, E.~I., \& 
Goldreich, P.\ 1997, \apj, 490, 368 

\bibitem[Dullemond \& Turolla(2000)]{dullemond2000} Dullemond, C.~P., 
\& Turolla, R.\ 2000, \aap, 360, 1187 

\bibitem[Dullemond \& Dominik(2004)]{dullemond2004} Dullemond, C.~P., 
\& Dominik, C.\ 2004, \aap, 417, 159 

\bibitem[Evans et al.(2003)]{evans2003} Evans, N.~J., II, et al.\ 
2003, \pasp, 115, 965 

\bibitem[Greene et al.(1994)]{greene1994} Greene, T.~P., Wilking, 
B.~A., Andre, P., Young, E.~T., \& Lada, C.~J.\ 1994, \apj, 434, 614 

\bibitem[Harvey et al.(2007)]{harvey2007} Harvey, P., Merin, B., 
Huard, T.~L., Rebull, L.~M., Chapman, N., Evans, N.~J., II, \& Myers, 
P.~C.\ 2007, ArXiv e-prints, 704, arXiv:0704.0009 

\bibitem[J{\o}rgensen et al.(2006)]{jorgensen2006} J{\o}rgensen, 
J.~K., et al.\ 2006, \apj, 645, 1246 

\bibitem[Kessler-Silacci et al.(2005)]{kessler2005} Kessler-Silacci, J.~E., 
Hillenbrand, L.~A., Blake, G.~A., \& Meyer, M.~R.\ 2005, \apj, 622, 404 

\bibitem[Lucy(1999)]{lucy1999} Lucy, L.~B.\ 1999, \aap, 344, 282 

\bibitem[Myers \& Ladd(1993)]{myers1993} Myers, P.~C., \& Ladd, 
E.~F.\ 1993, \apjl, 413, L47 

\bibitem[Pontoppidan et al.(in prep.)]{pontoppidan2007} Pontoppidan, K.~M., et al. 2007, in prep.

\bibitem[Porras et al.(2007)]{porras2007} Porras, A., et al.\ 
2007, \apj, 656, 493 

\bibitem[Robitaille et al.(2006)]{robitaille2006} Robitaille, T.~P., 
Whitney, B.~A., Indebetouw, R., Wood, K., \& Denzmore, P.\ 2006, \apjs, 
167, 256 

\bibitem[Sandford \& Allamandola(1993)]{sandford1993} Sandford, 
S.~A., \& Allamandola, L.~J.\ 1993, Icarus, 106, 478 

\bibitem[Ulrich(1976)]{ulrich1976} Ulrich, R.~K.\ 1976, \apj, 210, 377 

\bibitem[Weingartner \& Draine(2001)]{weingartner2001} Weingartner, 
J.~C., \& Draine, B.~T.\ 2001, \apj, 548, 296 

\bibitem[Whitney et al.(2003a)]{whitney2003a} Whitney, B.~A., Wood, 
K., Bjorkman, J.~E., \& Cohen, M.\ 2003a, \apj, 598, 1079 

\bibitem[Whitney et al.(2003b)]{whitney2003b} Whitney, B.~A., Wood, 
K., Bjorkman, J.~E., \& Wolff, M.~J.\ 2003b, \apj, 591, 1049 

\bibitem[Whitney et al.(2004)]{whitney2004} Whitney, B.~A., 
Indebetouw, R., Bjorkman, J.~E., \& Wood, K.\ 2004, \apj, 617, 1177 







\end{thebibliography}
\end{document}